\documentclass[3p,times,procedia]{elsarticle}
\usepackage{nupha_ecrc}
\usepackage{subfig}

\volume{00}

\firstpage{1}

\journalname{Nuclear Physics A}

\runauth{J.-F. Paquet}

\jid{nupha}

\jnltitlelogo{Nuclear Physics A}

\usepackage{amssymb}

\usepackage[figuresright]{rotating}

\begin{document}

\begin{frontmatter}



\dochead{XXVIth International Conference on Ultrarelativistic Nucleus-Nucleus Collisions\\ (Quark Matter 2017)}

\title{Probing the spacetime evolution of heavy ion collisions with photons and dileptons}


\author{Jean-Fran\c{c}ois Paquet}

\address{Department of Physics \& Astronomy, Stony Brook University, Stony Brook, NY 11794, USA}

\begin{abstract}
Photons and dileptons are emitted throughout the evolution of the deconfined nuclear medium produced in heavy ion collisions. As such they can provide valuable information about the different phases of the medium, and complement hadronic measurements and other observables. In this work, recent developments related to electromagnetic emissions at early time, in the cross-over region, and at late times are reviewed. The spacetime description of the nuclear medium from a hydrodynamic model of heavy ion collisions is used to provide context and guide the discussion.
\end{abstract}

\begin{keyword}
heavy-ion collisions \sep electromagnetic probes \sep viscous relativistic hydrodynamics \sep quark-gluon plasma

\end{keyword}

\end{frontmatter}


\section{Introduction}
\label{intro}

Heavy ion collisions at the Relativistic Heavy Ion Collider (RHIC) and the Large Hadron Collider (LHC) are used to produce an extended medium of deconfined nuclear matter often referred to as the ``quark-gluon plasma''. Before recombining into hadrons, on a timescale of order $10$~fm, this deconfined medium evolves through different phases which probes Quantum Chromodynamics (QCD) in a range of many-body settings.
There is considerable evidence that the deconfined matter is strongly coupled for most of this $\mathcal{O}(10~\textrm{fm})$ evolution~\cite{Shuryak:2004cy}. 
The spacetime expansion of the energy density and flow velocity of this nuclear medium has been modelled with viscous relativistic hydrodynamics from $\tau\sim 0.1~\textrm{fm}$ to $\sim10~\textrm{fm}$ (see e.g. Refs.~\cite{Gale:2013da,Heinz:2013th,deSouza:2015ena}). Such hydrodynamic models of heavy ion collisions have been successful in describing a wide array of soft hadron measurements with a comparatively limited set of parameters\footnote{
Assumptions about the underlying physics of heavy ion collisions are generally used to reduce considerably the number of free parameters in hydrodynamic models. If the medium is assumed to be in thermal and chemical equilibrium for most of its evolution, the equation of state is constrained by lattice calculations~\cite{Philipsen:2012nu}. Microscopic theories of early time dynamics can be used to estimate the macroscopic energy density and flow velocity of the medium at early time (see e.g. Section 4.1 of Ref.~\cite{deSouza:2015ena} for a recent overview). Relations between transport coefficients can be obtained in certain limits, and help reduce the number of independent coefficients to just a few (e.g. Ref.~\cite{Denicol:2014vaa}). Describing the late stage dynamics of the system with an afterburner helps reduce the dependence of observables on the details of the transition between hydrodynamics and late hadronic phase. Such assumptions can be relaxed and tested, moving the discussion toward physics rather than parameter counting.}. Ever more accurate and varied measurements of soft hadrons are providing new challenges and increasing constraints on hydrodynamic descriptions of heavy ion collisions. At the same time, theoretical developments are helping reduce the uncertainty of these same models, for example through an improved understanding of the early time dynamics of the medium~\cite{WilkeProc} and of the matching to hydrodynamics~\cite{Keegan:2016cpi, AleksasProc}.

The usual interpretation of viscous hydrodynamics is that it describes a fluid that is close to local thermal equilibrium, with deviations from equilibrium characterised by viscosities. In this picture, strong interactions between the medium's constituents are understood to maintain local equilibrium despite the explosive expansion of the fluid produced in heavy ion collisions, allowing a hydrodynamic description to be usable for a significant fraction of the medium's lifetime.

\begin{figure}[tb]
	\centering
  \subfloat[]{\includegraphics[width=0.38\textwidth]{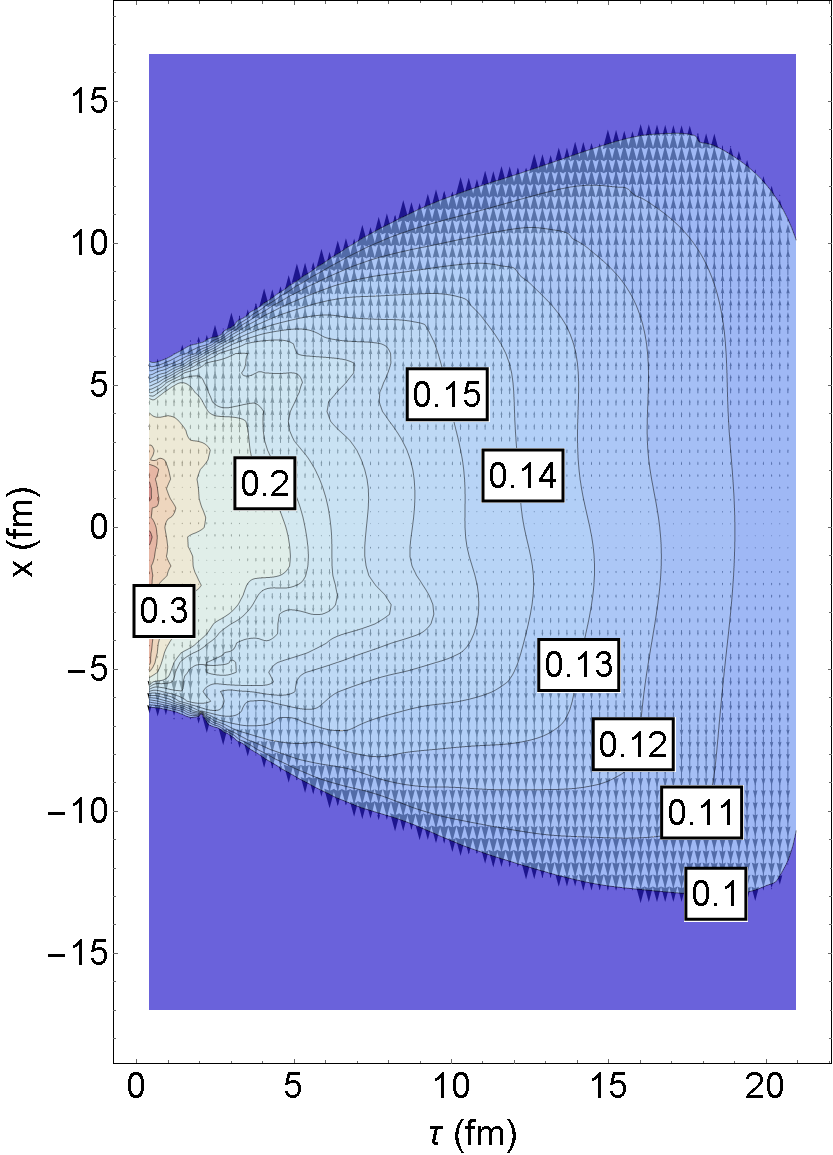}}%
	\hspace{0.3cm}
	\subfloat[]{\includegraphics[width=0.52\textwidth]{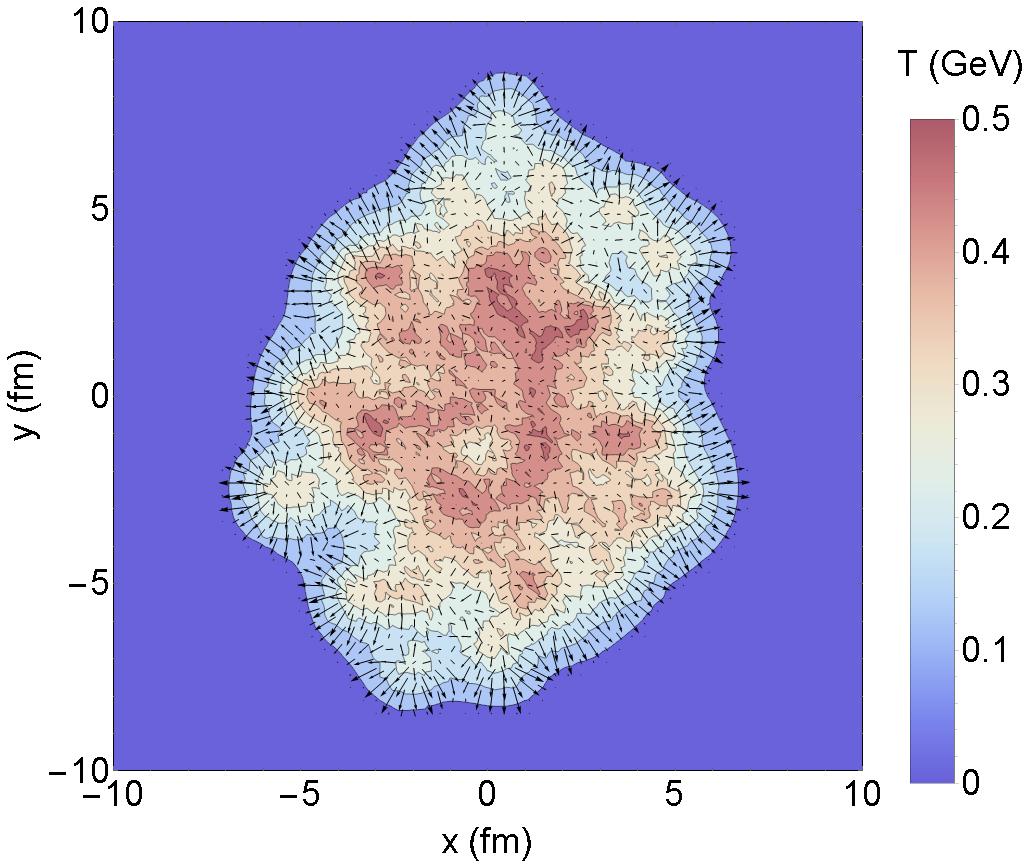}}%
	\caption{(a) Temperature profile (in GeV) at midrapidity from a hydrodynamic simulation for a central Au-Au collision with $\sqrt{s_{NN}}=200$~GeV. The abscissa is the time $\tau$ and the ordinate is a cut in the transverse plane. Contours are shown every 50 MeV for temperatures above 200 MeV, and every 10 MeV below. The arrows show the relative size of the flow velocity $u^\mu$ along the transverse axis. (b) Temperature and flow profile  in the transverse plane at $\tau_0=0.4$~fm. The same colour coding is used for both figures, but the arrows representing the flow velocity are scaled differently for clarity.}
	\label{fig:Tprofile}
\end{figure}

This interpretation of hydrodynamics as describing a fluid close to local equilibrium makes it possible to define a local temperature field across the region of applicability of hydrodynamics. This temperature is obtained from the energy density profile of hydrodynamics, through the lattice-constrained QCD equation of state~\cite{Philipsen:2012nu}. Figure~\ref{fig:Tprofile}(a) is an example of temperature profile obtained from a hydrodynamic simulation~\cite{Ryu:2015vwa} of a single Au-Au collision at the RHIC, for $\sqrt{s_{NN}}=200$~GeV. The profile is shown at midrapidity as a function of time $\tau$, for a cut in the transverse plane\footnote{Since this is a central event, a similar profile would be obtained for most choice of axis in transverse plane.} of the medium. The different contours indicate the temperature of the plasma in GeV. Overlaid with the temperature are arrows representing the relative size of the flow velocity $u^\mu$ along the chosen transverse direction.

The applicability of hydrodynamics is unclear once the medium is confined, since the interactions of hadrons may not be sufficiently strong to maintain local equilibrium. Confinement occurs progressively in a range of temperatures of order $150-200$~MeV for an equilibrated QCD medium without net baryon density. In regard of Fig.~\ref{fig:Tprofile}(a), it means that the temperature and flow profile significantly below $T\sim 150$~MeV should interpreted with care, and may only be rough estimates of the medium's profile at low temperatures.

Hydrodynamic simulations of heavy ion collisions assumes almost universally that hydrodynamics still provides a reasonable description of the medium at least slightly below confinement, such that when hydrodynamics is stopped, the medium can be described with hadronic degrees of freedom  (a process sometimes referred to as ``particlization''). 
This system of hadrons is still interacting; for comparisons with measurements, hadronic observables should be evaluated after these hadronic interactions become negligible. Nevertheless the importance of these final hadronic interactions varies from observable to observable. A number of important hadronic observables, such as the multiplicity, average transverse momentum and momentum anisotropy of charged hadrons or pions, are known not to be very sensitive to these interactions (see e.g. Ref.~\cite{Ryu:2017qzn} and references therein). These observables are thus dominated by the state of the medium at ``particlization''. If the properties of the medium at particlization is the result of the calculable hydrodynamic expansion at earlier times, soft hadronic observables can serve as indirect probes of the high temperature regions of the QCD medium.

Electromagnetic probes are one of many observables that are being investigated in heavy ion collisions to provide more direct information about the deconfined phase of the medium. Since the nuclear medium is small --- $\mathcal{O}(10~\textrm{fm})$ --- and the electromagnetic interaction is weak, photons and dileptons produced at any point in the medium can escape and be picked up by detectors. The deconfined medium is electrically charged and radiates photons and dileptons throughout its expansion. The momentum distribution of these electromagnetic probes reflect the local properties of the QCD medium at their point of emission, providing more direct information about the properties of the QCD medium at high temperature, and complementing soft hadrons observables.

Of the multiple sources of photons and dileptons in heavy ion collisions, hadronic decays (e.g. $\pi^0 \to \gamma \gamma$, $\omega \to \pi^0 e^+ e^-$) constitute a considerable background that often mask more interesting sources. For photons, hadronic decays are at least an order of magnitude larger than any other sources. Hadronic decays are also a significant background for dileptons in certain regions of momenta and invariant mass. This hadronic background, as well as other experimental challenges, mean that measurements of electromagnetic observables are currently limited in number. Moreover very few measurements have been made in a same system by two different experiments. As such, measurements of electromagnetic observables are still being actively investigated~\cite{CampbellProc}.


Besides this hadronic decay background, photons and dileptons are produced through multiple other mechanisms at different points during the collisions. It is convenient to look at these sources separately, starting with the early time ($\tau \lesssim 0.1-1$~fm) phase.

\section{Early times}
\label{sec:early}

Hydrodynamical models describe the evolution of the QCD medium produced in heavy ion collisions starting from a time $\tau \sim 0.1-1$~fm. Initial times of this order are used such that sufficient transverse flow is developed in the following $\mathcal{O}(10~\textrm{fm})$ of hydrodynamic evolution to describe measurements of the average transverse momentum and azimuthal momentum anisotropy ($v_n$) of charged hadrons. The initial time used in Fig.~\ref{fig:Tprofile}(a) is $0.4$~fm; the initial temperature and flow profile of the IP-Glasma initial conditions~\cite{Schenke:2012wb} used for this simulation is shown in Fig.~\ref{fig:Tprofile}(b).

Whether hadronic measurements can be described without such an early on-set of hydrodynamics evolution is not fully clear. This question was investigated from a phenomenological point of view in e.g. Ref.~\cite{Liu:2015nwa}. In parallel there is an on-going effort by the community to understand the theoretical foundation of equilibration and ``hydrodynamisation'' in heavy ion collisions. A general microscopic description of the early (pre-hydrodynamics) phase of heavy ion collisions has yet to be achieved, but considerable progress is being made~\cite{WilkeProc}.

Photon and dilepton observables are sensitive to the early phase of heavy ion collisions and as such are in a privileged position to contribute to the discussion. The effect of a rapid or delayed onset of parton chemical equilibrum on both photons and dileptons has been studied recently in hydrodynamic models~\cite{Vovchenko:2016ijt,Monnai:2014kqa}. Multiple investigations of electromagnetic emissions at early times have also been made in the past year~\cite{Greif:2016jeb,Berges:2017eom,Oliva:2017pri,TanjiProc}. It will be of great interest to see if these different approaches can come to a consensus regarding early chemical and thermal equilibration, as well as the amount of electromagnetic emissions produced during the first $\tau \sim 0.1-1$~fm of evolution of the deconfined medium. Comparisons with other non-equilibrium descriptions of heavy ion collisions~\cite{Linnyk:2015rco} would also be enlightening. Evaluating the azimuthal momentum anisotropies of this early radiation will be an important step as well to understand how they contribute to the photon $v_n$.

To help comparisons among different calculations and comparisons with data, it is important for studies of electromagnetic emission at early time to take full advantage of any constraints provided by hadronic observables. The total energy of the system at early time is constrained to a great extent by the hadron multiplicities measured at late times. Moreover any constraints on the flow velocity distribution at early times that can be obtained from measurements of hadronic average transverse momenta and azimuthal momentum anisotropies should be used.

A different source of photons and dileptons produced at early times originate from hard parton collisions: ``prompt photons'' and Drell-Yan dileptons~\cite{Ruuskanen:1991au}. In the range of dielectron invariant masses that are typically of interest to study the deconfined medium in heavy ion collisions ($0.3$~GeV$ \lesssim M_{e^+e^-} \lesssim 1$~GeV), Drell-Yan dileptons are smaller than other sources and are generally neglected. The situation is different for photons. At transverse momenta above $\sim 4-5$~GeV, prompt photons are the single dominant contribution to photon measurements once hadronic decays are subtracted (see e.g.~\cite{Adler:2005ig, Chatrchyan:2012vq}). At these high values of transverse momenta, prompt photons are predominantly produced in processes such as Compton scattering ($q g \to q \gamma$) and quark annihilation ($q \bar{q} \to g \gamma$), where the produced photon is the final state particle of a low level Feynman diagram. Prompt photons produced through such mechanisms are not affected by the final state parton energy loss encountered in heavy ion collisions. 

At transverse momenta below $\sim 4-5$~GeV, prompt photons are increasingly dominated by ``fragmentation photons'' --- soft photon radiated off perturbatively from a final state parton or produced non-perturbatively at the parton's fragmentation. These photons are affected by parton energy loss, and their evaluation demands a full simulation of energy loss in heavy ion collisions\footnote{Fragmentation photons are a subleading contribution to prompt photons at higher transverse momenta, but are not necessarily negligible. Estimates of the effect of energy loss on high transverse momenta photons can be found in e.g. Ref.~\cite{Arleo:2011gc}. It would be interesting to see if uncertainties on photon measurements can be reduced sufficiently to observe this fragmentation photon energy loss with high transverse momenta photons.}. Such calculations of prompt photons with parton energy loss have been made at the RHIC previously (e.g. Ref.~\cite{Turbide:2007mi}), but have yet to be undertaken at the LHC.


\section{Hydrodynamic evolution and confinement}
\label{sec:interm}

The early stage dynamics of the deconfined nuclear medium provides initial conditions for the hydrodynamics equations. From this point on, the evolution of the medium is determined by hydrodynamics, given an equation of state and transport coefficient of the medium constrained using theoretical calculations, models and hadronic observables. As seen in Figs.~\ref{fig:Tprofile}(a) and (b), the temperature and velocity profiles are lumpy and irregular at initial times, but becomes rapidly more uniform as the system evolves. The transverse expansion of the QCD medium stems primarily from the asymmetric energy distribution in the transverse plane of the early collisions, which originates both from the geometry of the collision and from initial state fluctuations. Asymmetry in the initial flow velocity is thought to play a lesser role. Hydrodynamics simulations of heavy ion collisions describe how the pressure gradients caused by these inhomogeneities in the initial energy develop into a significant flow velocity anisotropy at later time. This flow anisotropy is imprinted in the momentum of hadrons produced at particlization.

\begin{figure}[tb]
	\centering
  \subfloat[]{\includegraphics[width=0.41\textwidth]{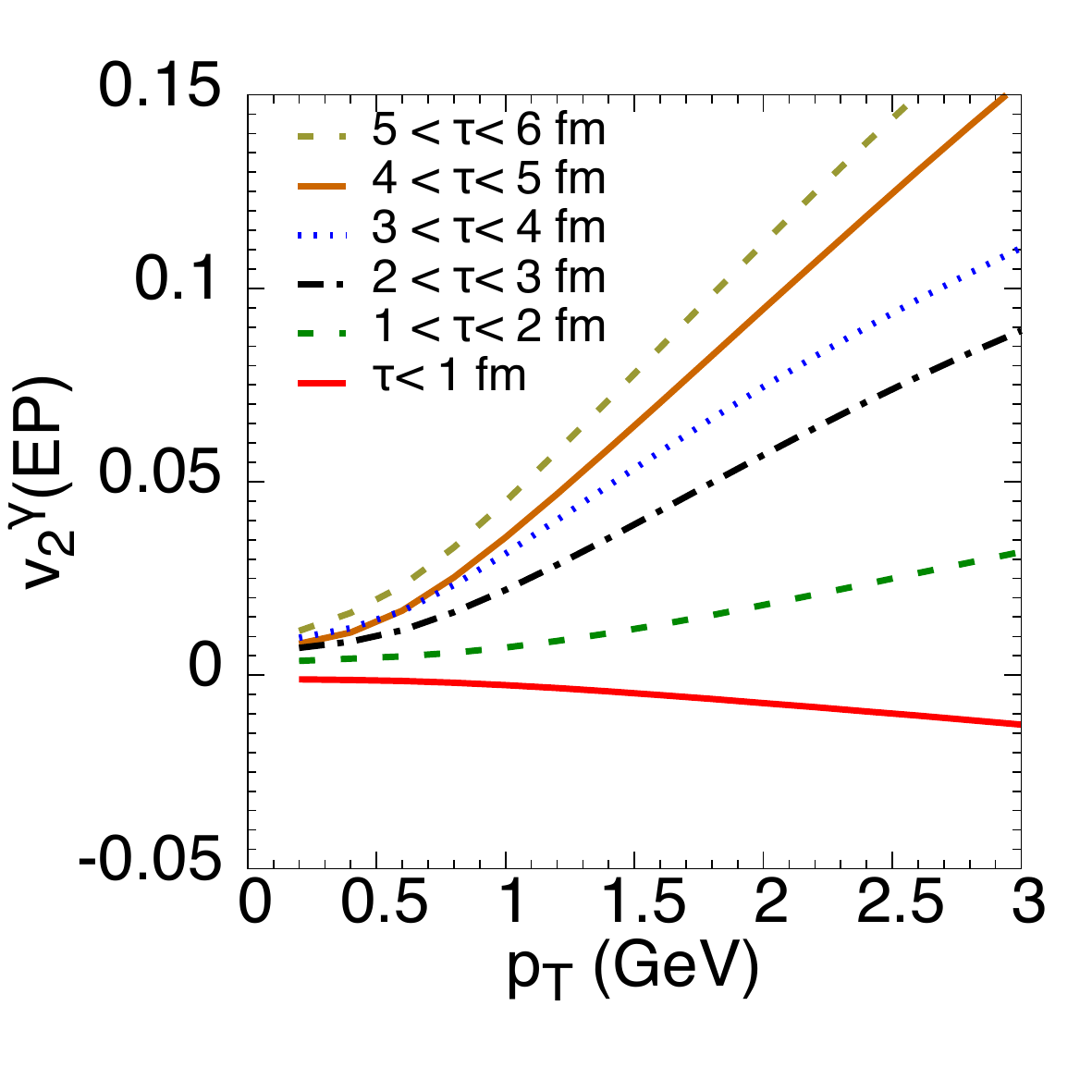}}%
	\hspace{0.3cm}
	\subfloat[]{\includegraphics[width=0.41\textwidth]{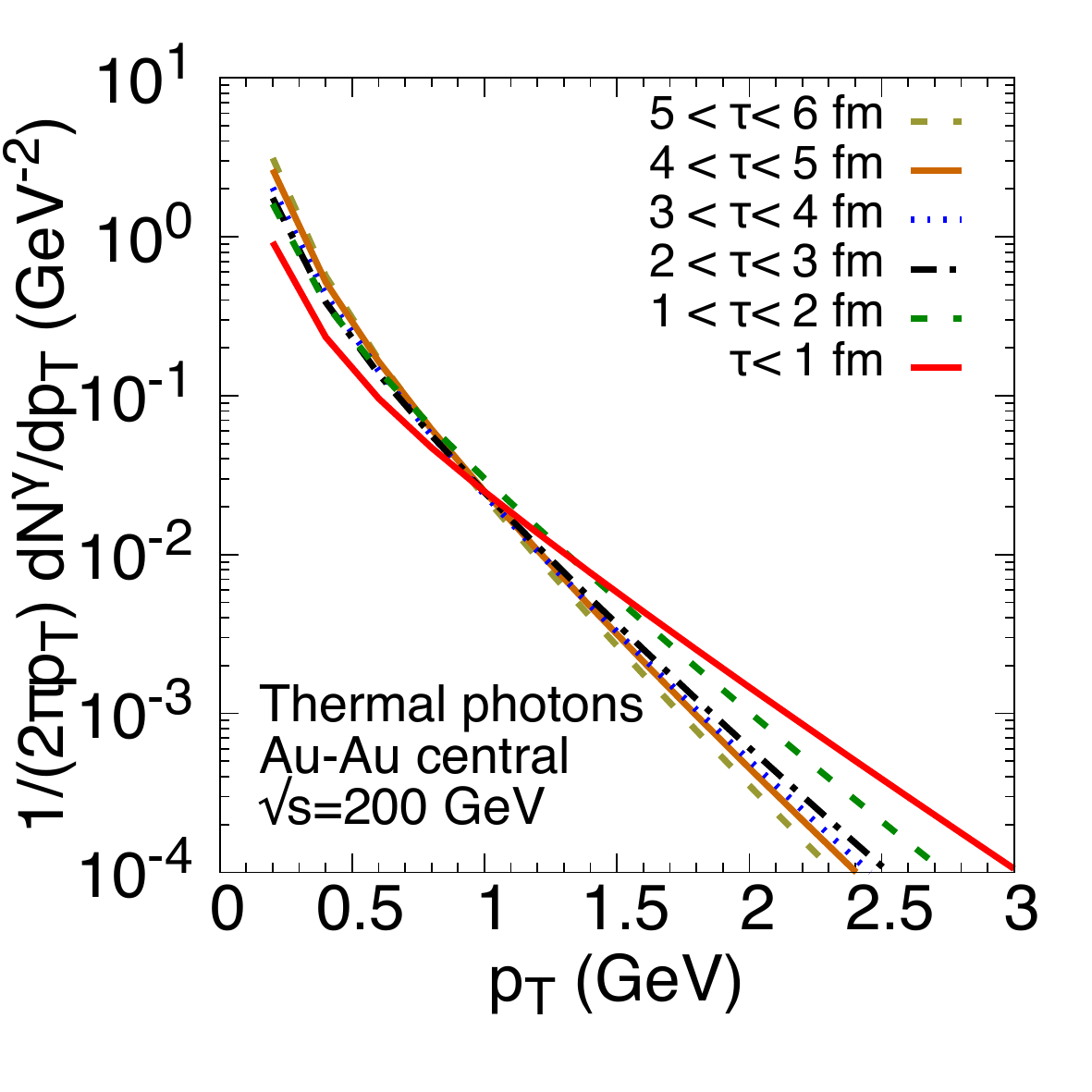}}%
	\caption{(a) Thermal photon $v_2(EP)$ and (b) thermal photon spectra at different times, for the central event illustrated in Fig.~\ref{fig:Tprofile}.}
	\label{fig:photonflow}
\end{figure}

Photons and dileptons produced by the medium are affected by this anisotropic expansion of the medium in the same way as hadrons, with the difference that they are emitted continuously. In this sense, electromagnetic emissions provide snapshots of the flow anisotropy of the medium throughout its evolution, although this information is largely averaged over in measurements. The thermal photon $v^\gamma_2(EP)$ at different times is shown in Fig.~\ref{fig:photonflow}(a) for the central Au-Au collision illustrated in Fig.~\ref{fig:Tprofile}. The photon  $v^\gamma_2(EP)$ is calculated with
\begin{equation}
v_n^\gamma(EP)=v_n^\gamma \cos(n(\Psi_n^\gamma-\Psi_n^h));\;\;\; v_n e^{i n \Psi_n}=\frac{\int d\phi (dN/d\phi) e^{i n \phi}}{\int d\phi (dN/d\phi)}
\end{equation}
with $dN/d\phi$ the photon, dilepton or hadron momentum distribution in azimuthal angle $\phi$.
As can be seen from Fig.~\ref{fig:photonflow}(a), photons have a small\footnote{In this specific instance, the $v^\gamma_2(EP)$ is small and negative at early time. The sign of the early $v^\gamma_2(EP)$ will differ from collision to collision, depending on the initial energy and flow profile, and the correlation between the photonic event plane $\Psi_2^\gamma$ at early time and hadronic event plane $\Psi_2^h$ at later times.} $v^\gamma_2(EP)$ early on ($\tau < 1$~fm), but this $v^\gamma_2(EP)$ grows rapidly with time. The $v^\gamma_2(EP)$ at times later than $\tau \sim 5-6$~fm are not shown, but are not significantly larger than the $v^\gamma_2(EP)$ in the $5 \lesssim \tau \lesssim 6$~fm range. That is, the flow anisotropy is already roughly saturated after $\sim 5$~fm of evolution.

The spectra of thermal photons is shown in Fig.~\ref{fig:photonflow}(b) for the same time intervals as Fig.~\ref{fig:photonflow}(a). The \emph{thermal} photon signal at large transverse momenta can be seen to be dominated by very early emissions ($\tau \lesssim 2$~fm), for reasons that are well-documented in the literature (see e.g. Ref.~\cite{Shen:2013vja}): as the medium expands, the temperature drops and the flow velocity increases. The temperature and the flow compete: the thermal photon spectra becomes softer with decreasing temperature, but harder with increasing flow velocity. At early times, the temperature decreases more rapidly than the flow increases, which translates into a rapid softening of the thermal photon spectra. After a few fermi of evolution, the flow is sufficiently large to make up for the loss of temperature, and the softening of the thermal photon spectra nearly stops. This can be seen in the similar slopes of the spectra for $\tau \sim 2-6$~GeV. A thorough analysis of these effects can be found in Ref.~\cite{Shen:2013vja}. The interesting consequence of this interplay between temperature and flow velocity in thermal photon production is that only very early thermal emission have a distinctive transverse momenta dependence in their spectra --- later time emissions have almost indistinguishable spectra.

By $\tau \sim 2-3$~GeV, most of the nuclear medium is below a temperature of $300$~MeV, and by $\tau \sim 5-6$~GeV --- the latest time shown in Fig.~\ref{fig:photonflow} --- most of it is below $200$~MeV. A large part of the thermal emission at the RHIC, even in this central collision, is thus produced at fairly low temperatures. A challenge in this range of temperature is to obtain the thermal electromagnetic emission rate from first principles. Recall that thermal electromagnetic probes are evaluated as
\begin{equation}
\frac{d^4 N_{\gamma/l^+l^-}}{d^4 K}= \int_{V_4} d^4 X \frac{d^4 \Gamma_{\gamma/l^+l^-}}{d^4 K}(K, u, T, \ldots)
\label{eq:thermal}
\end{equation}
where $V_4$ is the spacetime volume with local temperature $T(X) \gtrsim 100-150$~MeV (roughly speaking, the region that is not in deep blue in Fig.~\ref{fig:Tprofile}), and $d^4 \Gamma_{\gamma/l^+l^-}/d k^4$ is the thermal electromagnetic emission rate. The emission rate depends on the photon or dilepton four-momentum $K$, the flow velocity  $u$ and the temperature $T$. Corrections due to deviations from equilibrium (viscosity) should also be included as necessary\footnote{Note that viscous corrections to the photon emission rate are not included in the present work.}. 

At asymptotically high temperature, when the strong coupling constant $g_s$ is small, the emission rate $d^4 \Gamma_{\gamma/l^+l^-}/d k^4$ can be evaluated perturbatively~\cite{Arnold:2001ms,Aurenche:2002wq,Ghiglieri:2013gia,Ghiglieri:2014kma,Ghisoiu:2014mha,Ghiglieri:2015nba}. At low temperatures ($\sim100-150$~MeV), effective hadronic models can be used, e.g.~\cite{Turbide:2003si,Rapp:2000pe}. For phenomenologically relevant temperatures between $150$ and $300-400$~MeV, the electromagnetic rate is more challenging to evaluate: the nuclear medium is deconfined, but the strong coupling constant is not small. It is still common practice to estimate the electromagnetic rate in this range of temperatures by taking  the electromagnetic rates obtained through a perturbative expansion in $g_s$, and extrapolating it to a large value of the strong coupling constant ($g_s\sim 2$). This was the strategy used to make Fig.~\ref{fig:photonflow} (see Ref.~\cite{Paquet:2015lta} for more information about the rates used). However multiple different approaches have been used recently to investigate the photon and dilepton emission rates in this intermediate range of temperatures: comparisons with lattice results in the quenched limit~\cite{Ghiglieri:2016tvj}, holographic calculations~\cite{Finazzo:2015xwa,Iatrakis:2016ugz} and effective models of QCD at moderate temperatures~\cite{Gale:2014dfa,Lee:2014pwa}. Mixed results were obtained and a consensus has yet to emerge from these different calculations as to the value of the electromagnetic emission rates at moderate temperatures.

There have been suggestions that additional sources of electromagnetic emission could be present as the QCD medium goes through confinement (temperatures of order $150-200$~MeV)~\cite{Kharzeev:2013wra, Campbell:2015jga, Young:2015adw,ItakuraProc}. For example, descriptions of confinement as the recombination of effective quarks and gluons degrees of freedom into hadrons can have electromagnetic emissions as by-products. Since these photons would be emitted at late times, they would have a significant $v_2$, offering a compelling explanation for the large photon $v_2$ measured in heavy ion collisions. 

There are certain challenges in reconciling this picture with the current hydrodynamic descriptions of heavy ion collisions. As explained in the introduction, most simulations of heavy ion collisions assume that the medium can be described with hydrodynamics during the confinement phase
--- an assumption that appear to provide a good description of hadronic measurements.
If the medium is indeed close to local equilibrium through confinement, any electromagnetic emission from the medium should be reducible\footnote{Interactions of a non-thermal particle, a jet for example, with a thermal one is a separate case that cannot be included in the thermal emission rate and must be considered separately.} to a contribution to the thermal emission rate $d^4 \Gamma_{\gamma/l^+l^-}/d k^4$ (see Eq.~\ref{eq:thermal}). A large thermal rate around confinement is certainly favoured by current photon measurements~\cite{vanHees:2014ida}, but evaluating the thermal rate in this range of temperatures is still very challenging, as highlighted above; attempting to add electromagnetic emissions from specific mechanisms on top of a thermal emission rate that is still under investigation is difficult, and there is a risk of doublecounting emission sources. Such attempts must proceed with great care.

Should the QCD medium be significantly out-of-equilibrium during confinement, non-thermal sources of electromagnetic radiation would be present. Such scenarios require an alternative description of the medium in the spacetime volume where confinement is occurring, a description that is difficult, because the medium is strongly interacting and the degrees of freedom involved in this temperature range are unclear. Using the wealth of available hadronic observables to validate such non-equilibrium description of the confining medium is an important step to provide support for calculations of electromagnetic emission made within the same framework.

\section{Late times: electromagnetic emission below confinement}
\label{sec:late}

After $\mathcal{O}(10~\textrm{fm})$ of evolution, the QCD medium produced in heavy ion collisions is re-confined into an interacting system of hadrons. In the event that hadronic interactions can maintain local equilibrium,  photon and dilepton emissions can still be calculated using thermal emission rates with necessary corrections to account for viscosity. At some point, it should also be possible to describe this late stage of the medium with a kinetic theory description. Such a microscopic simulation of photon and dilepton emissions from interacting hadrons is challenging, as it involves describing the cross-section of every species of hadrons along with their electromagnetic radiation~\cite{Linnyk:2015rco,Bauchle:2010ym}. It is however of great interest to understand the progressive transition from hydrodynamics to particles~\cite{GojkoProc}, especially in view of the large photon $v_2$ measured at the RHIC, which is suggestive of a significant photon emission at late times.

Understanding this late stage of heavy ion collisions becomes increasingly important for small collision systems and lower center-of-mass energy collisions, where less of the spacetime evolution of the medium is expected to be described by hydrodynamics and more of it with kinetic theory. In this sense, a transport model that include electromagnetic emission from hadronic interactions will provide an important bridge between measurements of photons and dileptons at low energies (e.g. SPS, HADES, RHIC beam energy scan, and eventually FAIR) and at high energies (RHIC, LHC)~\cite{CampbellProc}. This is a crucial ingredient toward a systematic description of electromagnetic emission over a wide range of collision energies and system sizes. 

\section{Summary}

Electromagnetic probes are closely related to major challenges encountered in the study of heavy ion collisions: the onset of thermal and of chemical equilibrium at early time in the deconfined nuclear medium, the rapid development of flow velocity, the nature of the medium through the QCD cross-over and the transition to hadronic degrees of freedom at late times, to name only a few. Photon and dilepton emission can complement other observables by providing information about the different stages of evolution of the collisions. 

Multiple recent theoretical developments have been highlighted in this work. The number and scope of these studies promise significant advances in our understanding of electromagnetic emissions in heavy ion collisions for the near future. Combined with on-going investigations of electromagnetic observables by multiple collaborations at both the RHIC and the LHC~\cite{CampbellProc}, electromagnetic probes can be expected to contribute actively to our evolving understanding of heavy ion collisions and many-body QCD.

\paragraph*{Acknowledgements}

I thank Chun Shen for his valuable feedback on this work, and Jacopo Ghiglieri, Moritz Greif, Kazunori Itakura, Shu Lin, Naoto Tanji and Raju Venugopalan for their assistance in understanding and clarifying their recent works on photon production. I thank Charles Gale along with my colleagues in the nuclear theory group and the PHENIX photon group in Stony Brook for discussions and feedback that inspired part of this work. I thank Gojko Vujanovic for his help with dilepton-related topics. This work was supported by the U.S. D.O.E. Office of Science, under Award No. DE-FG02-88ER40388. Computations were made in part on the supercomputer Guillimin from McGill University, managed by Calcul Qu\'ebec and Compute Canada. The operation of this supercomputer is funded by the Canada Foundation for Innovation (CFI), Minist\`ere de l'\'Economie, des Sciences et de l'Innovation du Qu\'ebec (MESI) and le Fonds de recherche du Qu\'ebec – Nature et technologies (FRQ-NT).





\bibliographystyle{elsarticle-num}
\bibliography{biblio}







\end{document}